%% file: sample631.tex
\documentclass[]{aastex631}

\begin{document}

\title{Resolving the Galactic Center Region's Gamma-Ray Emission and Searching for its Neutrino Counterpart with HAWC and IceCube}

\input{HAWC-authors.tex}
\collaboration{91}{HAWC Collaboration}
\input{IceCube-authors.tex}
\collaboration{419}{IceCube Collaboration}

\correspondingauthor{Sohyoun Yun-C\'arcamo, Rishi Babu, IceCube Collaboration}
\email{sy645@drexel.edu, baburish@msu.edu, analysis@icecube.wisc.edu}



\begin{abstract} 
The Galactic Center (GC) is a prime candidate for petaelectronvolt cosmic-ray acceleration, with gamma-ray emission observed by the High Altitude Water Cherenkov (HAWC) Observatory extending beyond 100~TeV. Using $\sim$2860 days of HAWC data with improved machine-learning gamma/hadron separation, we present the first spatially resolved model of the GC region at the highest gamma-ray energies. The emission is resolved into a point source, HAWC J1745-2857, and a Galactic diffuse emission component that was not significant in previous HAWC analyses. With the diffuse emission accounted for, the point-source spectrum extends beyond 100~TeV with no evidence of a cutoff, showing that the PeVatron interpretation of the GC does not arise from unresolved diffuse emission. Assuming a hadronic origin, we construct the corresponding neutrino signal template and search for it in 12 years of IceCube muon-track data. No significant neutrino excess is found, and we report the first upper limit on the high-energy neutrino flux associated with the multi-component spatial model of the GC derived from HAWC data. The limit is approximately 12 times the nominal neutrino flux predicted from the HAWC gamma-ray model under a hadronic interpretation. Although the limit does not yet test the optically thin hadronic scenario, it quantifies the sensitivity required to test the hadronic origin of the GC PeVatron with future neutrino observations.

\end{abstract}



\section{Introduction} \label{sec:intro} 

The sources responsible for the observed Galactic cosmic-ray flux at petaelectronvolt (PeV) energies remain largely unidentified, representing a central open question in high-energy astrophysics \citep{aharonian2019massive,2021Univ....7..324C,2023PhRvD.107d3002S,2024NatAs...8..425D,blasi2013origin,amato2014origin,gabici2019origin}. While gamma-ray observations have identified several candidate PeVatrons \citep{2021NatAs...5..465A,LS5039,2024Natur.634..557A,2021ApJ...919L..22C}, these detections alone do not unambiguously establish hadronic acceleration, since both leptonic and hadronic processes can contribute to the observed emission. Complementary neutrino observations provide a powerful probe of the origin of high-energy cosmic rays. Gamma rays trace energetic particle interactions, while neutrinos are uniquely sensitive to hadronic processes through the decay of charged pions produced in cosmic-ray collisions. Therefore, the detection of both gamma rays and neutrinos from the same region would provide strong evidence for hadronic cosmic-ray acceleration \citep{2014JHEAp...1....1A,breuhaus2022galactic}.

The Galactic Center (GC) is one of the most promising targets for such studies. The High Altitude Water Cherenkov (HAWC) Observatory detected gamma rays from the GC extending beyond 100~TeV without evidence for a cutoff, providing strong evidence for proton acceleration to PeV energies \citep{hawcgc}. HAWC currently remains the only observatory capable of monitoring the GC above $\sim$40~TeV, making it uniquely sensitive to the highest-energy gamma-ray emission from this region. Concurrently, the IceCube Neutrino Observatory has measured an astrophysical neutrino flux from the Galactic plane \citep{IceCube:2023ame}, motivating targeted searches for spatially correlated emission from candidate sources such as the GC.

In this \textit{Letter}, we present an updated analysis of the GC region using an extended HAWC dataset with improved event discrimination \citep{Alfaro_2025} compared to previous studies. In contrast to the aforementioned HAWC analysis, which modeled the emission with a single point source \citep{hawcgc}, and in which a diffuse component was tested but was not significant, the present study benefits from increased exposure and improved background rejection, enabling a spatially resolved characterization of both diffuse and point-source components in the GC region. This is the first spatially resolved model of the GC region at the highest gamma-ray energies, and we test whether the PeVatron spectrum of the central source persists once the diffuse emission is accounted for. The obtained best-fit gamma-ray model is then used to predict the expected neutrino emission, assuming a fully hadronic origin for the point-source emission and using only the pion-decay component of the diffuse emission. Using this template, we perform a search for neutrino emission with IceCube's muon track data, enabling a direct test of hadronic scenarios for the GC region and placing constraints on the associated high-energy neutrino flux. Previous IceCube analyses constrained hadronic emission from the Galactic plane as a whole using large-scale diffuse templates \citep{Aartsen_2017,Aartsen_2019,IceCube:2023ame}; this work provides the first IceCube test of a joint diffuse and point-source template around the GC.

This \textit{Letter} is organized as follows. Section~\ref{sec:data} describes the HAWC and IceCube datasets used in this analysis. Section~\ref{sec:methods} details the gamma-ray modeling, neutrino template construction, and likelihood analysis methods. Section~\ref{sec:main} presents the results of the HAWC analysis and IceCube neutrino constraints. Finally, Section~\ref{sec:conc} summarizes the implications of these results for the origin of Galactic cosmic rays.

\section{Data}\label{sec:data}

\subsection{HAWC data} 

The HAWC Observatory is located on the slope of the Sierra Negra volcano in Puebla, Mexico, at 4100 m above sea level. The main detector consists of 300 water Cherenkov detectors, which measure Cherenkov light from extensive air showers initiated by gamma and cosmic rays interacting with atmospheric nuclei \citep{ABEYSEKARA2023168253}. Air-shower reconstruction is performed using the ``Pass 5'' algorithms detailed in \cite{albert2024performance}, while including the updated machine-learning technique \citep{Alfaro_2025} to separate the gamma-ray signal from the dominant cosmic-ray background ($\sim$99.9\% of events). Relative to the earlier Pass~4 reconstruction, Pass~5 improves the shower core and direction reconstruction and recovers high-energy events that were previously rejected, increasing HAWC's sensitivity at the highest energies \citep{albert2024performance}.

We use $\sim$2860 days of data analyzed with the Pass 5 neural-network (NN) energy estimator \citep{albert2024performance}. As in the previous publication, the dataset includes off-array events, defined as showers with cores that fall outside the main array by up to 1.5 times its physical area, which enhance HAWC’s sensitivity at high zenith angles and energies \citep{albert2024performance}. This is particularly relevant for the GC, which culminates at a zenith angle of $\sim$48$^{\circ}$ at the HAWC site \citep{Alfaro_2025}. Compared with the dataset used in the previous HAWC GC analysis \citep{hawcgc}, the present dataset includes approximately one additional year of exposure and adopts the machine-learning gamma/hadron separation of \citet{Alfaro_2025}, which increases the detection significance of the GC by $\sim$34\% relative to the standard selection. The energy range constrained by this analysis is determined by the fit and is shown in Figure~\ref{fig:hawc_gc_flux_2}.

\subsection{IceCube data} 

The IceCube Neutrino Observatory instruments a cubic kilometer of Antarctic ice at the Amundsen-Scott South Pole Station. Its 5160 digital optical modules, deployed across 86 vertical strings, detect the Cherenkov light produced when neutrinos interact with the ice \citep{2017JInst..12P3012A}. The observed light patterns depend on the neutrino flavor and interaction type. Muon neutrinos undergoing charged-current interactions either within the instrumented volume or in the surrounding ice produce muon tracks. These tracks, produced predominantly by the outgoing muon in $\nu_{\mu}$ charged-current interactions, are straight and can extend for many kilometers, allowing detection even when the interaction vertex lies outside the detector. This geometry provides excellent angular resolution, making the dataset ideal for point-source searches \citep{Abbasi_2025}. The choice of a track sample for the extended template used in this analysis is motivated in Section~\ref{sec:neutrino}. The GC lies in the southern sky; therefore, searches for neutrino emission from this region rely on down-going track events (as seen by IceCube), which are contaminated primarily by atmospheric muons. This background is reduced using event-selection criteria based on reconstructed event properties, including energy and reconstruction quality, together with boosted decision tree classifiers \citep{2020PhRvL.124e1103A}. In this analysis, we use 12 years of data from the point-source tracks (PSTracks) selection, collected with the full 86-string IceCube configuration, and exclude partial-detector configurations (IC40--IC79). The event selection follows \citet{2020PhRvL.124e1103A}. Although the IceCube Upgrade has recently expanded the detector with more densely instrumented strings in the central DeepCore region, this analysis does not include Upgrade data.


\section{Methodology}\label{sec:methods} 

Our analysis proceeds in two stages. We first determine the best-fit gamma-ray emission model for the GC region using HAWC data. This model is then used to construct the corresponding neutrino emission template to be fit using IceCube data.

\subsection{HAWC gamma-ray modeling}\label{sec:gammamod} 

We model the gamma-ray flux from the GC region using the HAWC Accelerated Likelihood (HAL) plugin within the Multi-Mission Maximum Likelihood (3ML)\footnote{https://github.com/threeML/threeML} framework \citep{younk2015high,abeysekara2022characterizing}. The analysis employs a forward-folded maximum-likelihood method \citep{Vianello:201667}. 

The test statistic (TS$_{\text{HAWC}}$) is defined as
\begin{equation}\label{ts}
\text{TS}_{\text{HAWC}} = -2\ln{\left(\frac{L_{\text{bkg}}}{L_{\text{model}}}\right)}\,,
\end{equation}
where $L_{\text{model}}$ is the maximum likelihood of the source model and $L_{\text{bkg}}$ corresponds to the background-only hypothesis. Under Wilks’ theorem \citep{wilks1938large}, which is applicable to HAWC data \citep{Abeysekara_2017}, the TS$_{\text{HAWC}}$ follows a $\chi^{2}$ distribution with degrees of freedom equal to the difference in the number of free parameters between nested models. 

Initial source candidates within the region of interest (ROI) are identified using a source seed-finding procedure. This procedure is based on the image-based source-search pipeline \citep{drip} and is applied to the NN point-source significance map shown in Figure \ref{fig:hawc_gc_data}. We validate this approach using an independent blind multi-source search method \citep{2017ApJ...843..139A}, following previous HAWC analyses \citep[e.g.,][]{albert2024tevanalysissourcerich,alfaro2024testingmolecularcloudparadigm}. Both methods consistently identify a single point source, which may include contributions from HESS J1745-290 (Sgr A$^{*}$; \citealp{Aharonian_2006}) and HESS J1746-285 (the Arc; \citealp{abdalla2018characterising}). Their individual fluxes are small compared to the GC diffuse ridge flux \citep{hawcgc}, corresponding to the extended TeV gamma-ray emission surrounding Sgr A$^{*}$ that traces the local dense molecular clouds. Both power-law and curved spectral models are tested, with the power-law model providing the preferred description of the point-source spectrum.

We then test the inclusion of Galactic diffuse emission (GDE) using the High-Energy Radiative MESsengers (HERMES) framework \citep{hermes}, which accounts for inverse Compton scattering, Bremsstrahlung, and neutral pion ($\pi^0$) decay from cosmic-ray interactions with interstellar gas. The spatial templates for atomic hydrogen and $^{12}$CO distributions are taken from \citet{h14pi} and \citet{dame}, respectively, with $^{12}$CO used as a molecular gas tracer. Cosmic-ray proton and electron models in HERMES are adopted from \cite{fornieri2020features}. For the $\pi^0$ decay component, we adopt the cross-section parameterization from \citet{kelneraharonian}.

The preferred model combines the single point source with the GDE component. The free parameters are the point source position (right ascension and declination), flux normalization, spectral index, and a scaling factor for the GDE template.
The point-source spectrum is modeled as
\begin{equation}\label{power law}
\frac{\mathrm{d}N}{\mathrm{d}E} = \phi_{16} \left(\frac{E}{E_0}\right)^{-\gamma}\,,
\end{equation}
where $\phi_{16}$ is the flux normalization at the pivot energy $E_0 = 16$~TeV, and $\gamma$ is the spectral index. The pivot energy was calculated such that it minimizes the correlation between the flux normalization and spectral index. The GDE is scaled by the best-fit dimensionless normalization factor that multiplies the predicted gamma-ray intensity, which is defined per unit solid angle (sr$^{-1}$).

The best-fit fluxes of the point source and GDE at the GC position are then used to determine the relative normalization between the point-source and diffuse components for the construction of the neutrino template described in the following section.  Nearby unresolved or sub-threshold sources, such as SNR G0.9+0.1 and HESS J1745-303, were not identified as significant excesses in the source search described above and may therefore contribute to the diffuse component of the model, with minimal impact on the total emission compared to the central point source.

\subsection{Neutrino template construction}\label{sec:neutrino}

Under the assumption of hadronic interactions, the same cosmic-ray population responsible for gamma-ray emission via $\pi^0$ decay also produces neutrinos through charged pion decay. We assume that the parent cosmic-ray interactions occur through $pp$ collisions only. Therefore, the neutrino spatial template is constructed from the HAWC gamma-ray best-fit model by selecting only the hadronic $\pi^0$-decay component of the GDE template. This component is approximately 10 times more dominant than the leptonic component and is used to define the spatial morphology and energy dependence of the diffuse neutrino emission. The neutrino flux is related to the gamma-ray flux produced in $pp$ interactions following the prescription from \citet{Ahlers_2014}.

Then, we incorporate a point-source component located at the GC. The point-source spectral index and the diffuse-to-point-source flux ratio at the GC pixel are fixed to the values obtained from the HAWC gamma-ray fit. The resulting HAWC-motivated neutrino template is therefore defined by the spatial morphology and energy dependence of the HERMES $\pi^0$-decay GDE component together with the GC point-source component, with its spectral index and relative normalization fixed by the HAWC best-fit model. The neutrino and antineutrino fluxes are combined, and neutrino oscillations are assumed to result in an approximately flavor-equal flux at Earth.

The resulting neutrino emission model is convolved with IceCube’s detector acceptance and each event’s angular uncertainty to construct an event-wise signal probability density function. The overall normalization of the combined neutrino template is parameterized solely by the total number of signal events, $n_s$. Throughout this work, the “model flux” refers to the nominal neutrino flux predicted by the HAWC-motivated template, obtained by summing the flux contribution from all spatial pixels of the combined GDE and point-source template.

Although the likelihood formalism follows previous IceCube extended-source template analyses, the HAWC-derived template is restricted to the $8^{\circ}\times8^{\circ}$ HAWC ROI centered at the GC. This was to reduce the contamination of nearby brighter gamma-ray sources and isolate the point-source component at the GC: the point source is well localized by HAWC, and at its position the diffuse emission contributes only $\sim$35\% of the point-source flux (Section~\ref{sec:HAWCres}). The neutrino template is therefore not a smooth diffuse template, but a well-localized point source embedded in diffuse emission. This motivates the use of track events, whose angular resolution of $\lesssim 1^{\circ}$ at TeV energies resolves the point source and the relative contributions of the point-source and diffuse components fixed by HAWC. Cascade events, with an angular resolution of around 5$^{\circ}$ at high energies \citep{IceCube2024showers}, comparable to the size of the entire region, cannot resolve the point source, so the two components could not be distinguished from each other or from the surrounding Galactic plane emission \citep{IceCube:2023ame}.

To avoid experimenter bias, the analysis is performed in a blinded manner, with the likelihood construction and validation carried out using right-ascension-scrambled data and simulations. Possible outcomes are studied a priori through signal injection and background trials, and the true source region is only examined after the analysis procedure is finalized. For extended sources, right-ascension scrambling can introduce signal contamination into the background estimation. The likelihood framework accounts for this effect through the same signal subtraction procedure and likelihood formalism used in previous IceCube extended-source template analyses \citep{Aartsen_2017, Aartsen_2019,IceCube:2023ame}. The GDE spectral shape, the point-source spectral index, and the diffuse-to-point-source normalization ratio derived at the GC pixel are fixed by the HAWC-motivated neutrino template described above and are not free parameters in the likelihood fit. Therefore, the only free parameter is the overall signal normalization, $n_s$. Consequently, the test statistic (TS$_{\text{IceCube}}$) is defined as
\begin{equation}
\text{TS}_{\text{IceCube}} = -2 \ln \left( \frac{\mathcal{L}(n_s = 0)}{\mathcal{L}(\hat{n}_s)} \right),
\end{equation}
where $\hat{n}_s$ is the best-fit number of signal events that maximizes the likelihood $\mathcal{L}(n_s)$ and $\mathcal{L}(n_s = 0)$ corresponds to the background-only hypothesis. The background distribution is estimated using 5,000,000 pseudo-experiments generated by randomizing each event's right ascension. The resulting best-fit signal normalization is converted into a flux normalization, expressed in units of the model-predicted flux.

\section{Results}\label{sec:main}

\subsection{HAWC best-fit results}\label{sec:HAWCres} 

\begin{figure}[ht!]
    \centering
    \includegraphics[width=0.5\columnwidth]{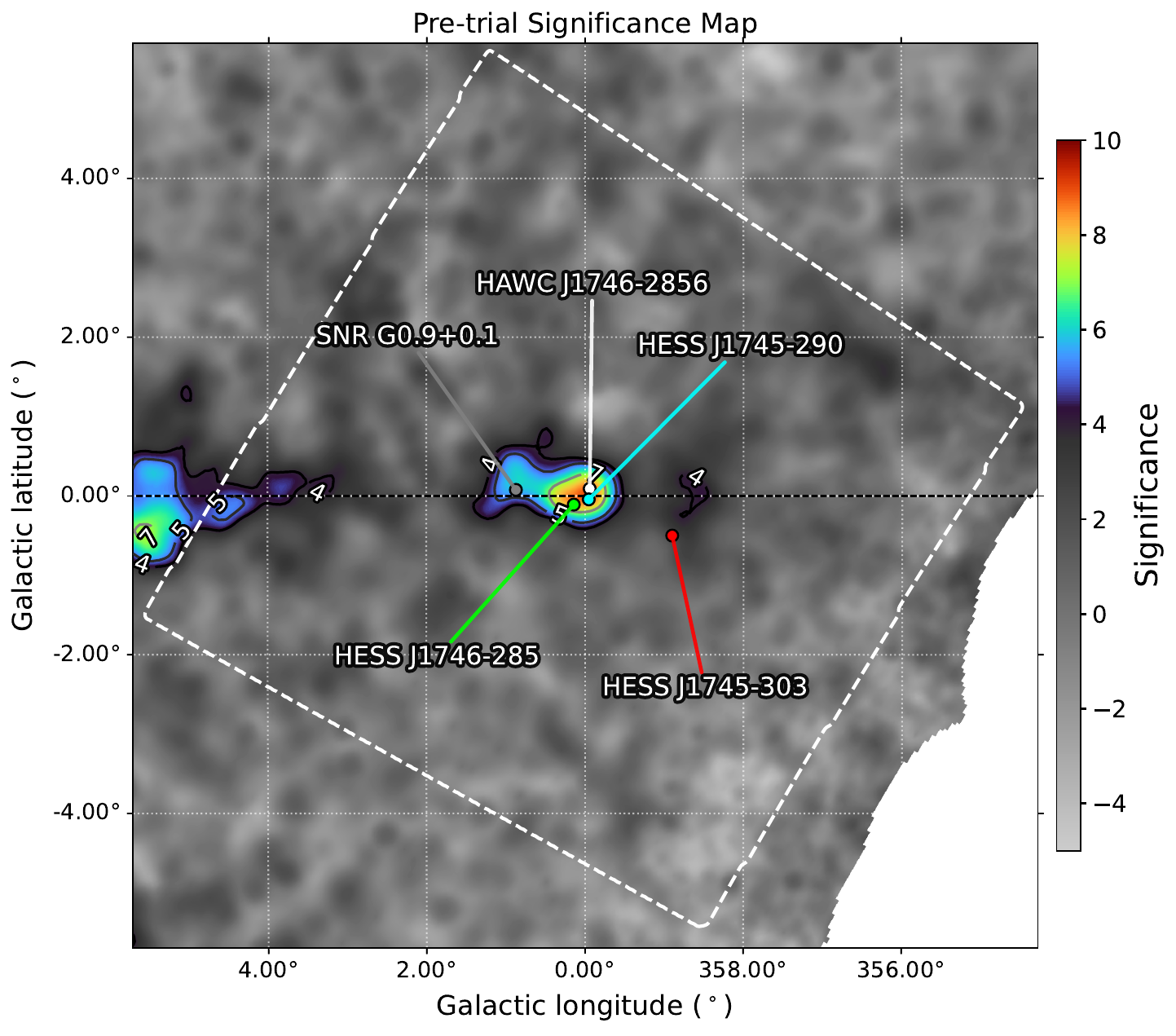}
    \caption{Pre-trial significance map of the GC region obtained using the HAWC NN energy estimator (including both on- and off-array events; \citealp{abeysekara2019measurement, albert2024performance}), derived under the assumption of a point-source morphology and a power-law spectrum with index 2.6. The map is shown in grayscale for significances below 4$\sigma$ and in color for higher values. The peak significance is 9$\sigma$. Contours, which are numbered in the plots, indicate levels of statistical significance. Colored markers with leader lines indicate the best-fit positions of known TeV sources, with colors used solely for visual identification: HAWC J1746-2856 (white; \citet{hawcgc}), HESS J1745-290 (Sgr A$^{*}$; cyan; \citet{Aharonian_2006}), HESS J1745-285 (the Arc; green; \citet{abdalla2018characterising}), SNR G0.9+0.1 (yellow; \citet{2005A&A...432L..25A}), and HESS J1745-303 (red; \citet{2018A&A...612A...1H}). The dashed white rectangle denotes the boundary of the ROI.  The white pixels at the bottom right region mark the edge of HAWC's field of view for the energy estimator maps. 
    }
    \label{fig:hawc_gc_data}
\end{figure}

HAWC detects the GC with a pre-trial peak significance of 9$\sigma$ above the expected background. The corresponding significance map is shown in Figure~\ref{fig:hawc_gc_data}. The ROI is a square of side $8^{\circ}$ centered on the Galactic coordinate origin, in equatorial coordinates. This region was chosen to encompass the GC and its surroundings without truncation at the edge of the HAWC field of view, as evidenced by the masked (white) pixels in the lower-right of Figure~\ref{fig:hawc_gc_data}. The choice of this ROI size allows the inclusion of events with relatively large angular uncertainties while reducing contamination from nearby bright gamma-ray sources. Smaller ROIs were tested to completely exclude the source at the northern edge of the ROI, and verified that excluding it does not significantly change the fit results.

The best-fit parameters with their respective statistical and systematic uncertainties are shown in Table~\ref{tab:sys} for both the point source and diffuse component. The systematic uncertainties account for non-negligible, independent systematic uncertainties identified in a previous energy-dependent study of the Crab Nebula \citep{abeysekara2019measurement}, which are evaluated by generating instrument response functions under different detector configurations to assess potential detector mis-modeling.

\begin{table}[h]
    \centering
    \begin{tabular}{lllrcc}
    \hline
    Component & TS$_{\text{HAWC}}$ & Parameter & Best fit & Statistical & Systematic \\
    & & & & uncertainties & uncertainties \\
    \hline
    \hline
    
    & & Right ascension \hfill ($^{\circ}$) & 266.33 & $\pm$0.06 & $\pm$0.10 \\
    HAWC J1745-2857& 35.5 & Declination \hfill ($^{\circ}$) & $-28.95$ & $\pm$0.05 & $\pm$0.09 \\
    & & $\phi_{16}$ ($10^{-15}$ \hfill TeV$^{-1}$ cm$^{-2}$ s$^{-1}$) & 4.90 & $\pm$0.91 & $\pm$1.54 \\
   & & Spectral index  & $2.99$ & $\pm$0.16 & $\pm$0.28 \\
    \hline
    HERMES GDE  & 69  & Scale factor \hfill (dimensionless) & 2.51 & $\pm$0.32 & $\pm$0.43 \\
    \hline
    \end{tabular}
    \caption{Best-fit parameters for the HAWC GC analysis including statistical and systematic uncertainties. The model consists of a point source (HAWC J1745-2857) and the GDE component modeled by a template generated with the HERMES software. The table reports results from separate likelihood fits for the point-source and diffuse components. The point-source parameters include RA, Dec, flux normalization at 16~TeV, and spectral index. The best fit position of HAWC J1745-2857 corresponds to $(l, b) = (-0.05^{\circ}, 0.05^{\circ})$ in Galactic coordinates. The GDE component is characterized by a single dimensionless scale factor multiplying the HERMES gamma-ray intensity template (per unit solid angle). The corresponding TS$_{\text{HAWC}}$ values are listed for each component fit.}
    \label{tab:sys}
\end{table}

\begin{figure}[ht!]
    \centering
    \includegraphics[width=0.48\columnwidth]{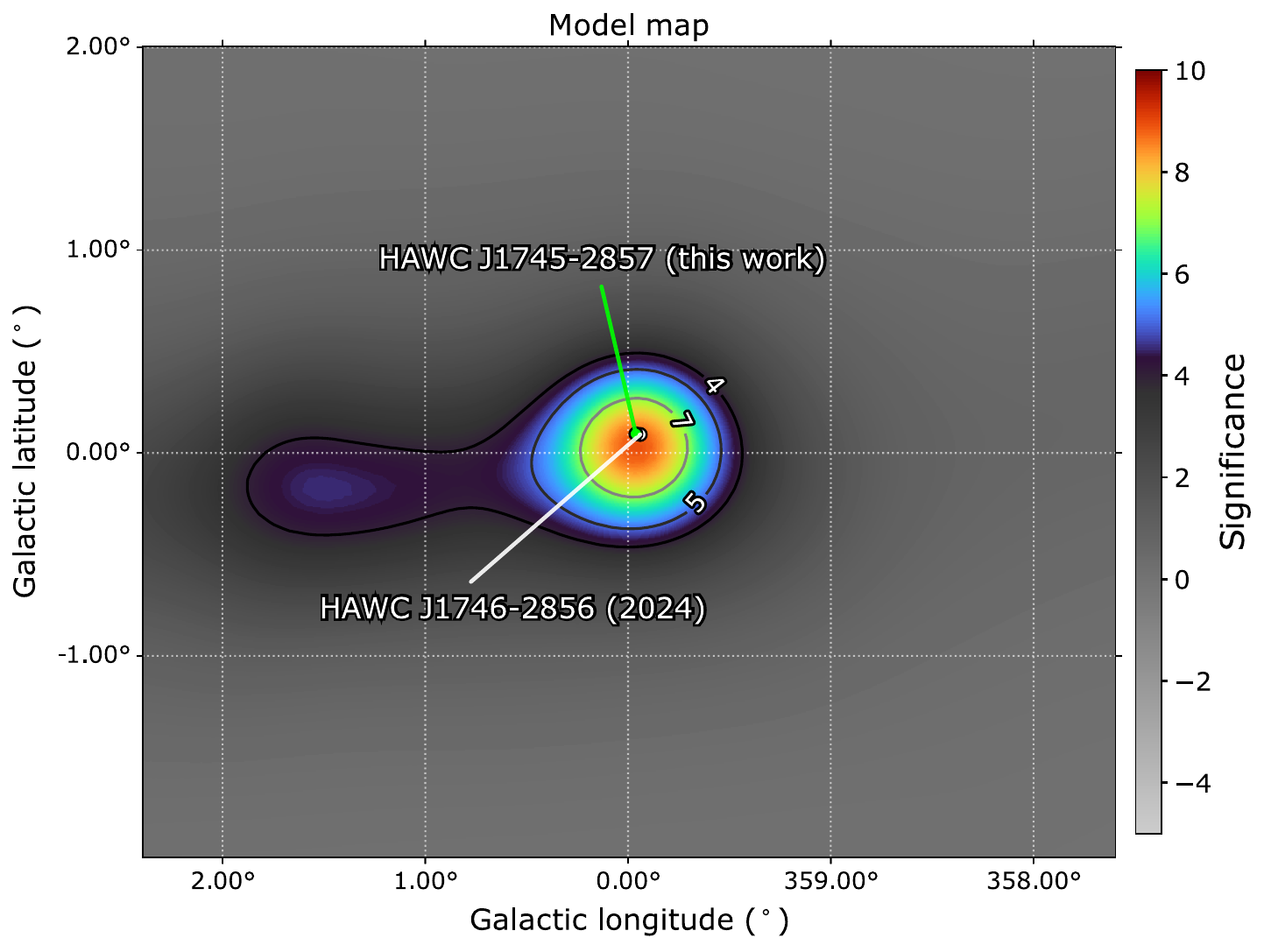}
    \includegraphics[width=0.48\columnwidth]{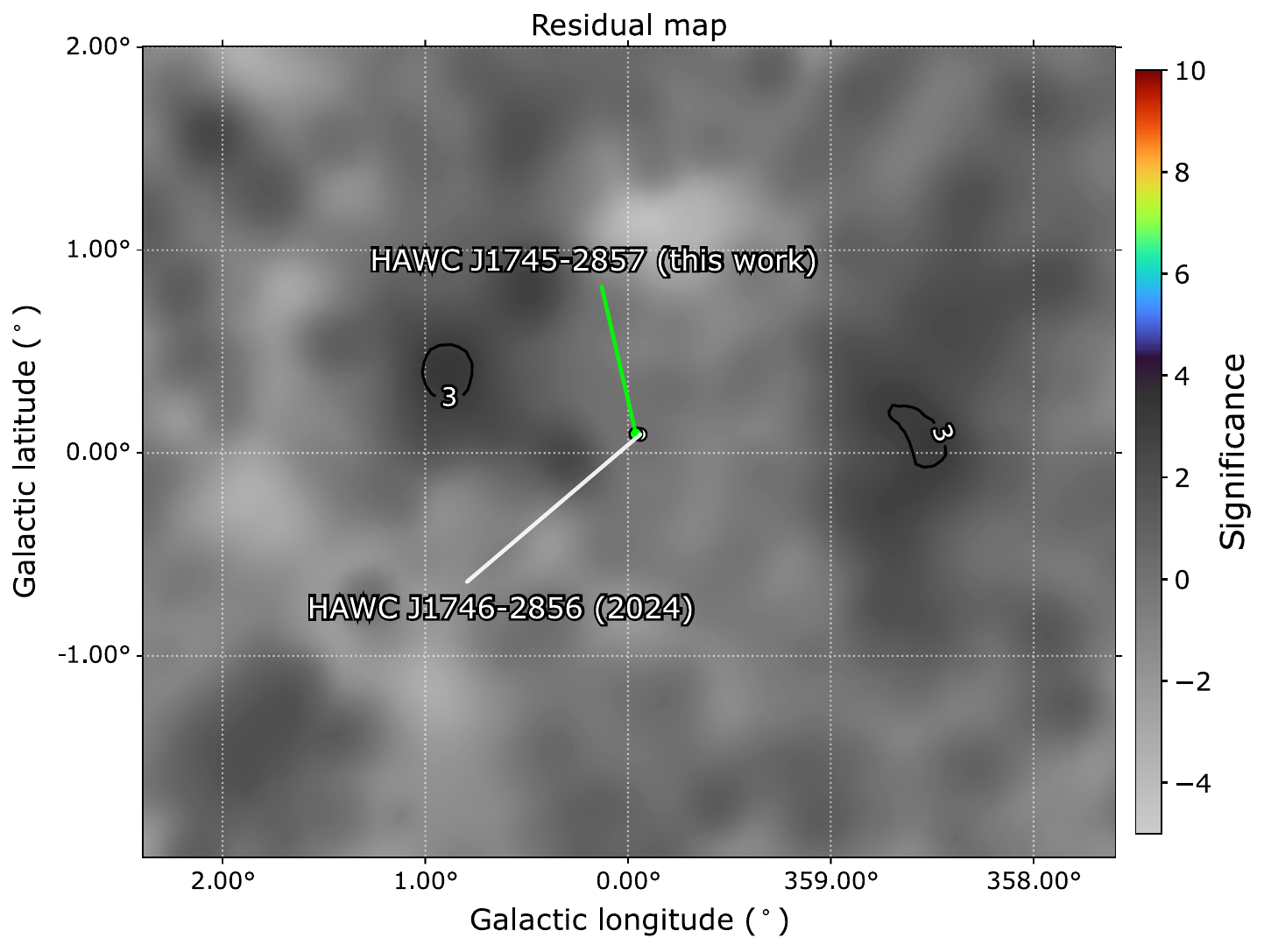}
    \caption{Post-fit maps of the GC region. The contours are labeled with their corresponding values to indicate levels of statistical significance. The map is shown in grayscale for significances below 4$\sigma$ and in color for higher values. Colored markers with leader lines indicate the best-fit positions of the point source for this work (green) and the previous HAWC publication (white), with the colors used solely for visual identification. \textbf{Left:} HAWC best-fit model map of the GC region including the point source and diffuse emission. \textbf{Right}: Residual map after subtracting the model from the data, showing no significant unmodeled features.}
    \label{fig:hawc_model_residuals}
\end{figure}

In Figure~\ref{fig:hawc_model_residuals}, we show the HAWC best-fit model significance map (left panel) and the corresponding residual map (right panel) obtained after subtracting the model from the data. These maps illustrate that the combination of point-source and diffuse emission provides a good description of the observed gamma-ray distribution in the GC region. The updated point-source localization differs from earlier HAWC results due primarily to the increased data set and improved reconstruction algorithms, and remains statistically consistent with the previous localization, both with and without the GDE model.

Figure~\ref{fig:hawc_gc_flux_2} (left panel) shows the best-fit gamma-ray spectrum of the GC point source obtained with the preferred model, which includes both point-source and GDE components, compared with the point-source-only fit. The point-source-only fit using the updated dataset is consistent with the previously published HAWC point-source-only result (not shown; \citealp{hawcgc}). Including the GDE component reduces the fitted point-source flux, as expected, with a diffuse-to-point-source flux ratio of $\sim$35\% at the GC position.

\subsection{IceCube upper limit} 

\begin{figure}[ht!]
    \centering
    \includegraphics[width=0.48\columnwidth]{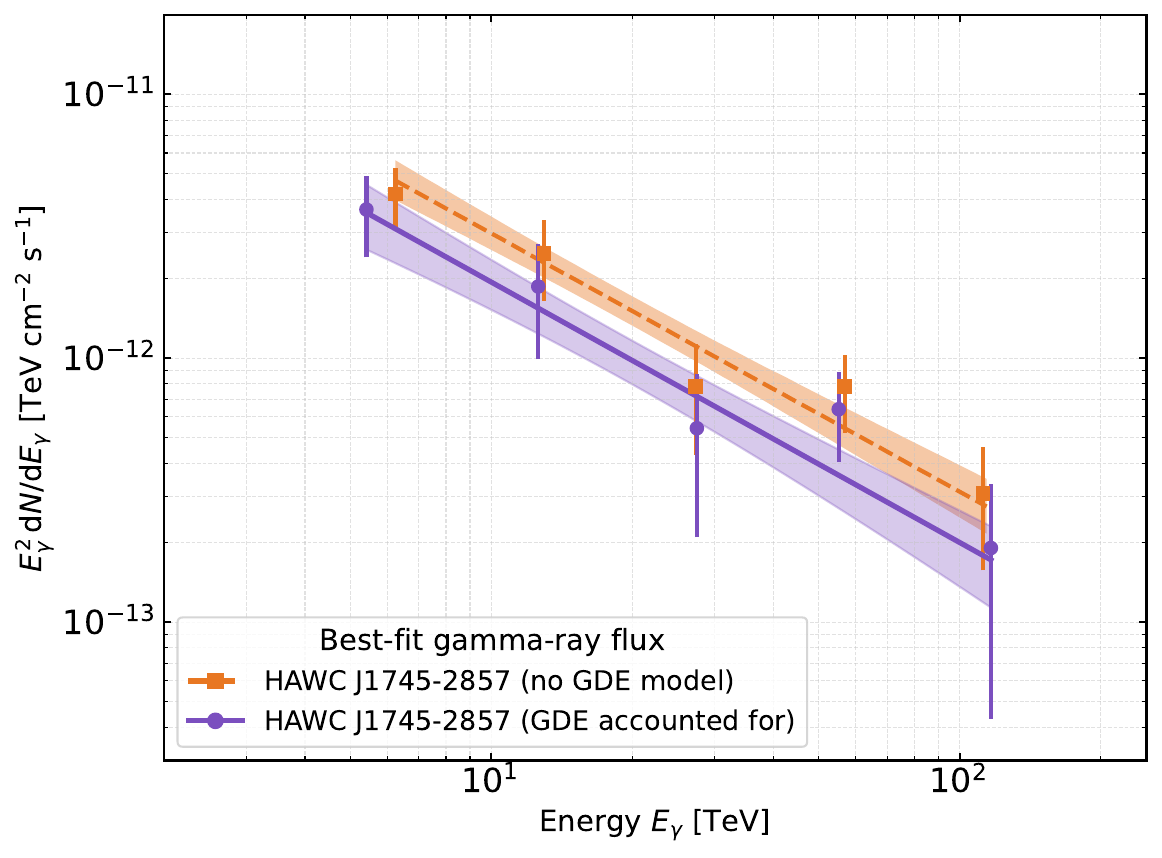}
    \includegraphics[width=0.48\columnwidth]{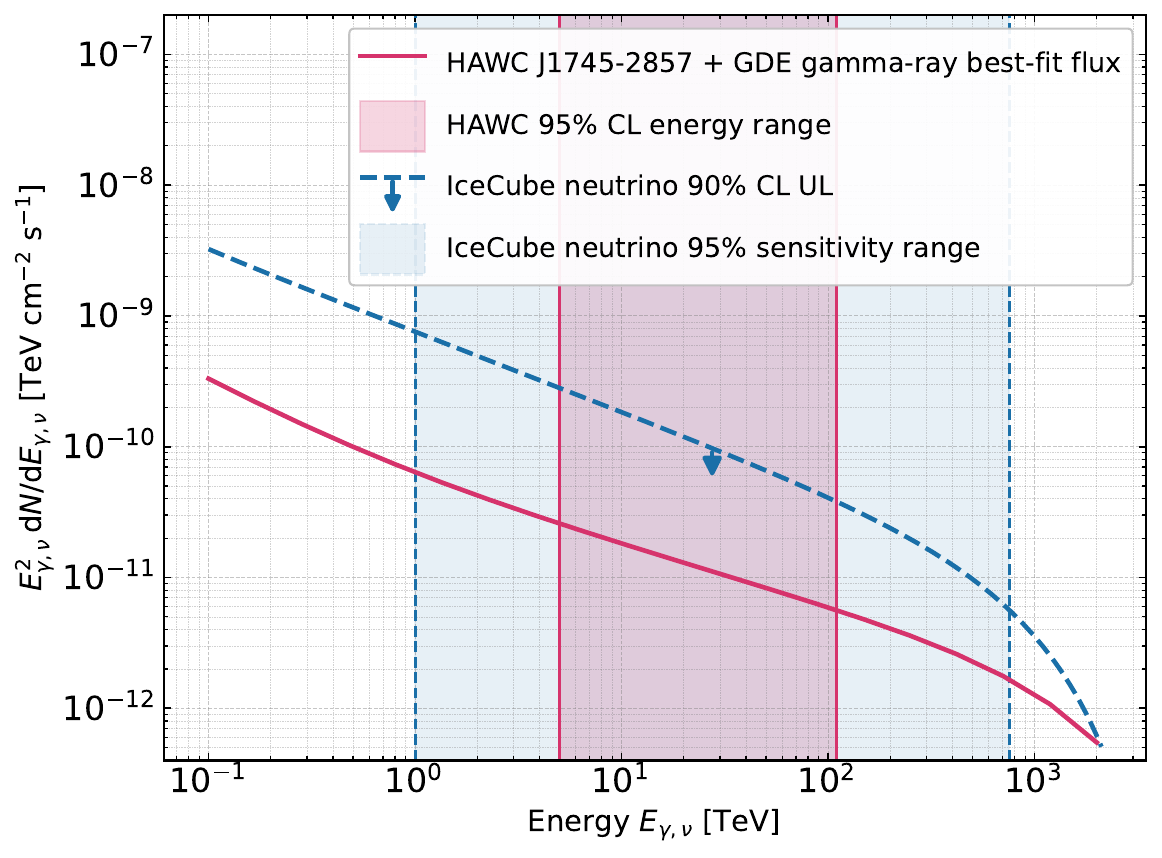}
    \caption{Best-fit results. \textbf{Left}: HAWC best-fit gamma-ray ($\gamma$) flux from the point-source component when the model also includes the GDE component (purple solid line and circle markers), compared with the point-source-only model (orange dashed line and square markers). \textbf{Right}: Neutrino ($\nu$) per-flavor UL derived using IceCube PSTracks (blue dashed line), together with the HAWC best-fit gamma-ray flux from the GDE template plus the point-source component (red solid line). Here, $\nu$ denotes the sum of neutrino and antineutrino fluxes for each flavor. The neutrino UL and gamma-ray best-fit flux correspond to the total scaled flux obtained by summing over all pixels in the respective templates. While the gamma-ray GDE model includes multiple emission channels (more details in Section~\ref{sec:gammamod}), the neutrino template is derived from the $\pi^{0}$ component of the GDE and the separately fitted point source, assuming a hadronic origin. The shaded regions indicate the 95\% CL energy interval of the HAWC model flux (red) and the 95\% energy range contributing most strongly to the IceCube PSTracks flux UL (blue).}
    \label{fig:hawc_gc_flux_2}
\end{figure}

The data show no preference for a non-zero signal normalization in the GC region, under the assumption of the model derived from the HAWC observations, consistent with a background-only hypothesis ($p$-value $=0.49$). The 3$\sigma$ evidence potential and 5$\sigma$ discovery potential for this analysis are 26 and 42 times the nominal HAWC-motivated neutrino model flux, respectively. Therefore, we place a 90\% confidence level (CL) upper limit (UL) on the normalization of the HAWC-motivated neutrino template, corresponding to approximately 12 times the nominal neutrino model flux. Here, the model flux refers to the neutrino flux predicted from the HAWC gamma-ray model under the assumed hadronic $pp$ scenario, rather than to the HAWC gamma-ray flux shown in Figure~\ref{fig:hawc_gc_flux_2}. We note that this analysis does not include corrections for gamma–gamma absorption, as the results are derived purely from the observed emission in the region.

Figure~\ref{fig:hawc_gc_flux_2} (right panel) shows the 90\% CL UL (blue dashed line) on the HAWC-motivated neutrino emission template, compared with the HAWC best-fit gamma-ray flux from the GDE template plus the point-source component (red solid line). The gamma-ray GDE model includes several emission components, while the neutrino template is constructed from the $\pi^{0}$-dominant component of the GDE together with the separately fitted point-source contribution, assuming a hadronic origin. The shaded bands indicate the 95\% CL energy range of the HAWC analysis (red) and the 95\% sensitivity range of IceCube PSTracks (blue).

\section{Conclusion}\label{sec:conc} 

We have performed an updated analysis of the GC region using $\sim$2860 days of HAWC data. The best-fit gamma-ray model is comprised of a point source (HAWC J1745-2857), potentially including contributions from Sgr A$^{*}$ and the Arc, together with a GDE component modeled using HERMES that may also absorb unresolved or sub-threshold sources. Including the HERMES diffuse template reduces the fitted point-source flux relative to a point-source-only fit, with the diffuse emission contributing $\sim$35\% of the point-source flux at the source location. The point source therefore still dominates the emission at the GC, while the diffuse component makes a significant contribution. Despite this additional diffuse component, the preferred spectral model and spectral index of the point source remain unchanged. 

Based on this HAWC gamma-ray model, we have constructed a gamma-ray-informed neutrino template comprising both the point-source and diffuse emission components inferred from the gamma-ray analysis, while fixing the relative flux normalization between the two components and the point-source spectral index to the best-fit HAWC values. Using this template, we have performed a likelihood-based search with 12 years of IceCube muon-track data. No significant neutrino excess is detected, and we have derived the first IceCube 90\% CL UL on the high-energy neutrino flux for this HAWC-informed, multi-component spatial model of the GC region, corresponding to approximately 12 times the nominal neutrino flux predicted from the HAWC gamma-ray model under a purely hadronic emission assumption. The UL therefore does not yet test the optically thin hadronic interpretation. Scenarios involving significant gamma-ray attenuation could produce larger neutrino-to-observed-gamma-ray ratios, but their interpretation is model dependent.  

The remaining sensitivity gap highlights the importance of improved neutrino reconstruction and the combination of complementary event samples. Combining through-going muon tracks with starting-track selections, such as the Enhanced Starting Track Event Selection (\citealp{ESTES}), can improve sensitivity to southern sky sources by increasing statistics while enhancing background rejection and energy reconstruction through contained interaction vertices. Future analyses combining different event selections and improved reconstruction techniques will be crucial for probing faint neutrino emission from Galactic PeVatrons.

Looking ahead, next-generation detectors such as IceCube-Gen2 \citep{aartsen2021icecube}, KM3NeT \citep{margiotta2014km3net}, and Pacific Ocean Neutrino Experiment (P-ONE; \citealp{bailly2021two}) will expand neutrino astronomy capabilities through increased exposure, improved event reconstruction, and complementary sky coverage. In particular, Northern Hemisphere observatories like KM3NeT and P-ONE will provide an Earth-filtered view of the GC, enhancing muon background rejection. Together, these advances bring sensitivities closer to model-predicted Galactic neutrino source fluxes and enable stronger tests of hadronic scenarios in the GC.

In summary, the gamma-ray analysis presents the first spatially resolved model of the GC region at the highest gamma-ray energies and reinforces the HAWC PeVatron interpretation of the GC point source by confirming a power-law spectrum extending beyond 100 TeV without evidence for a cutoff, while accounting for a spatially resolved GDE component, showing that the PeVatron interpretation of the GC does not arise from unresolved diffuse emission. Notably, this analysis explicitly includes and fits a spatially resolved GDE model in the GC region at energies above 40 TeV, extending previous very-high-energy studies of the region. We further present the first search for a neutrino counterpart to this HAWC GC observation including both diffuse and point-source emission components, which is also the first IceCube test of a joint diffuse and point-source template around the GC, and derive a corresponding neutrino UL that provides constraints for hadronic acceleration scenarios and future multi-messenger studies of the GC. 

\begin{acknowledgments}
The HAWC Collaboration acknowledges the support from: the US National Science Foundation (NSF); the US Department of Energy Office of High-Energy Physics; the Laboratory Directed Research and Development (LDRD) program of Los Alamos National Laboratory; Consejo Nacional de Ciencia y Tecnolog\'{i}a (CONACyT), M\'{e}xico, grants LNC-2023-117, 271051, 232656, 260378, 179588, 254964, 258865, 243290, 132197, A1-S-46288, A1-S-22784, CF-2023-I-645, CBF2023-2024-1630, c\'{a}tedras 873, 1563, 341, 323, Red HAWC, M\'{e}xico; DGAPA-UNAM grants IG101323, IG100726, IN111716-3, IN111419, IA102019, IN106521, IN114924, IN110521 , IN102223; VIEP-BUAP; PIFI 2012, 2013, PROFOCIE 2014, 2015; the University of Wisconsin Alumni Research Foundation; the Institute of Geophysics, Planetary Physics, and Signatures at Los Alamos National Laboratory; Polish Science Centre grant, 2024/53/B/ST9/02671; Coordinaci\'{o}n de la Investigaci\'{o}n Cient\'{i}fica de la Universidad Michoacana; Royal Society - Newton Advanced Fellowship 180385; Gobierno de Espa\~na and European Union-NextGenerationEU, grant CNS2023- 144099; The Program Management Unit for Human Resources \& Institutional Development, Research and Innovation, NXPO (grant number B16F630069); Coordinaci\'{o}n General Acad\'{e}mica e Innovaci\'{o}n (CGAI-UdeG), PRODEP-SEP UDG-CA-499; Institute of Cosmic Ray Research (ICRR), University of Tokyo. H.M. acknowledges support under grant number CBF2023-2024-1630. H.F. acknowledges support by NASA under award number 80GSFC21M0002. C.R. acknowledges support from National Research Foundation of Korea (RS-2023-00280210, RS-2026-25470489). We also acknowledge the significant contributions over many years of Stefan Westerhoff, Gaurang Yodh and Arnulfo Zepeda Dom\'inguez, all deceased members of the HAWC collaboration. Thanks to Scott Delay, Luciano D\'{i}az and Eduardo Murrieta for technical support.

The IceCube collaboration acknowledges the significant contributions to this manuscript from R. Babu and S. Yun-C\'arcamo. The IceCube authors gratefully acknowledge the support from the following agencies and institutions: USA – U.S. National Science Foundation-Office of Polar Programs, U.S. National Science Foundation-Physics Division, U.S. National Science Foundation-EPSCoR, U.S. National Science Foundation-Office of Advanced Cyberinfrastructure, Wisconsin Alumni Research Foundation, Center for High Throughput Computing (CHTC) at the University of Wisconsin–Madison, Open Science Grid (OSG), Partnership to Advance Throughput Computing (PATh), Advanced Cyberinfrastructure Coordination Ecosystem: Services \& Support (ACCESS), Frontera and Ranch computing project at the Texas Advanced Computing Center, U.S. Department of Energy-National Energy Research Scientific Computing Center, Particle astrophysics research computing center at the University of Maryland, Michigan State University, Astroparticle physics computational facility at Marquette University, NVIDIA Corporation, and Google Cloud Platform; Belgium – Funds for Scientific Research (FRS-FNRS and FWO), FWO Odysseus and Big Science programmes, and Belgian Federal Science Policy Office (Belspo); Germany – Bundesministerium für Forschung, Technologie und Raumfahrt (BMFTR), Deutsche Forschungsgemeinschaft (DFG), Helmholtz Alliance for Astroparticle Physics (HAP), Initiative and Networking Fund of the Helmholtz Association, Deutsches Elektronen Synchrotron (DESY), and High Performance Computing cluster of the RWTH Aachen; Sweden – Swedish Research Council, Swedish Polar Research Secretariat, Swedish National Infrastructure for Computing (SNIC), and Knut and Alice Wallenberg Foundation; European Union – EGI Advanced Computing for research; Australia – Australian Research Council; Canada – Natural Sciences and Engineering Research Council of Canada, Calcul Québec, Compute Ontario, Canada Foundation for Innovation, WestGrid, and Digital Research Alliance of Canada; Denmark – Villum Fonden, Carlsberg Foundation, and European Commission; New Zealand – Marsden Fund; Japan – Japan Society for Promotion of Science (JSPS), Ministry of Education, Culture, Sports, Science and Technology (MEXT), and Institute for Global Prominent Research (IGPR) of Chiba University; Korea – National Research Foundation of Korea (NRF); Switzerland – Swiss National Science Foundation (SNSF).
\end{acknowledgments}

\vspace{5mm}





\bibliography{sample631}{}
\bibliographystyle{aasjournal}



\end{document}

%% file: HAWC-authors.tex
\author{R.~Alfaro}
\affiliation{Instituto de F\'{i}sica, Universidad Nacional Autónoma de México, Ciudad de Mexico, Mexico }

\author[0000-0002-0595-9267]{M.~Araya}
\affiliation{Universidad de Costa Rica, San José 2060, Costa Rica}

\author{J.C.~Arteaga-Velázquez}
\affiliation{Universidad Michoacana de San Nicolás de Hidalgo, Morelia, Mexico }

\author[0000-0002-5529-6780]{R.~Babu}
\affiliation{Department of Physics and Astronomy, Michigan State University, East Lansing, MI, USA }

\author{P.~Bangale}
\affiliation{Temple University, Department of Physics, 1925 N. 12th Street, Philadelphia, PA 19122, USA}

\author{A.~Bernal}
\affiliation{Instituto de Astronom\'{i}a, Universidad Nacional Autónoma de México, Ciudad de México, México.}

\author[0000-0002-4042-3855]{K.S.~Caballero-Mora}
\affiliation{Universidad Autónoma de Chiapas, Tuxtla Gutiérrez, Chiapas, México}

\author[0000-0003-2158-2292]{T.~Capistrán}
\affiliation{Instituto Nacional de Astrof\'{i}sica, Óptica y Electrónica, Puebla, Mexico }

\author[0000-0002-8553-3302]{A.~Carramiñana}
\affiliation{Instituto Nacional de Astrof\'{i}sica, Óptica y Electrónica, Puebla, Mexico }

\author{F.~Carreón}
\affiliation{Instituto de Astronom\'{i}a, Universidad Nacional Autónoma de México, Ciudad de México, México.}

\author[0000-0002-6144-9122]{S.~Casanova}
\affiliation{Institute of Nuclear Physics Polish Academy of Sciences, PL-31342 IFJ-PAN, Krakow, Poland }

\author[0000-0002-7607-9582]{U.~Cotti}
\affiliation{Universidad Michoacana de San Nicolás de Hidalgo, Morelia, Mexico }

\author[0000-0002-1132-871X]{J.~Cotzomi}
\affiliation{Facultad de Ciencias F\'{i}sico Matemáticas, Benemérita Universidad Autónoma de Puebla, Puebla, Mexico }

\author[0000-0002-7747-754X]{S.~Coutiño de León}
\affiliation{Dept. of Physics and Wisconsin IceCube Particle Astrophysics Center, University of Wisconsin{\textemdash}Madison, Madison, WI, USA}

\author[0000-0002-8528-9573]{C.~de León}
\affiliation{Universidad Michoacana de San Nicolás de Hidalgo, Morelia, Mexico }

\author[0000-0001-9643-4134]{E.~De la Fuente}
\affiliation{Departamento de F\'{i}sica, Centro Universitario de Ciencias Exactase Ingenierias, Universidad de Guadalajara, Guadalajara, Mexico }

\author[0000-0002-7574-1298]{N.~Di Lalla}
\affiliation{Department of Physics, Stanford University: Stanford, CA 94305–4060, USA}

\author{R.~Diaz Hernandez}
\affiliation{Instituto Nacional de Astrof\'{i}sica, Óptica y Electrónica, Puebla, Mexico }

\author[0000-0002-0087-0693]{J.C.~Díaz-Vélez}
\affiliation{Dept. of Physics and Wisconsin IceCube Particle Astrophysics Center, University of Wisconsin{\textemdash}Madison, Madison, WI, USA}

\author[0000-0002-2987-9691]{M.A.~DuVernois}
\affiliation{Dept. of Physics and Wisconsin IceCube Particle Astrophysics Center, University of Wisconsin{\textemdash}Madison, Madison, WI, USA}

\author[0000-0001-5737-1820]{K.~Engel}
\affiliation{Department of Physics, University of Maryland, College Park, MD, USA }

\author[0000-0003-2423-4656]{T.~Ergin}
\affiliation{Department of Physics and Astronomy, Michigan State University, East Lansing, MI, USA }

\author[0000-0001-7074-1726]{C.~Espinoza}
\affiliation{Instituto de F\'{i}sica, Universidad Nacional Autónoma de México, Ciudad de Mexico, Mexico }

\author[0000-0002-0173-6453]{N.~Fraija}
\affiliation{Instituto de Astronom\'{i}a, Universidad Nacional Autónoma de México, Ciudad de México, México.}

\author[0000-0002-4188-5584]{J.A.~García-González}
\affiliation{Tecnologico de Monterrey, Escuela de Ingenier\'{i}a y Ciencias, Ave. Eugenio Garza Sada 2501, Monterrey, N.L., Mexico, 64849}

\author[0000-0003-1122-4168]{F.~Garfias}
\affiliation{Instituto de Astronom\'{i}a, Universidad Nacional Autónoma de México, Ciudad de México, México.}

\author{N.~Ghosh}
\affiliation{Department of Physics, Michigan Technological University, Houghton, MI, USA }

\author{S.~Giri}
\affiliation{Institute of Nuclear Physics Polish Academy of Sciences, PL-31342 IFJ-PAN, Krakow, Poland }

\author{A.~Gonzalez Muñoz}
\affiliation{Instituto de F\'{i}sica, Universidad Nacional Autónoma de México, Ciudad de Mexico, Mexico }

\author[0000-0002-5209-5641]{M.M.~González}
\affiliation{Instituto de Astronom\'{i}a, Universidad Nacional Autónoma de México, Ciudad de México, México.}

\author{J.A.~González}
\affiliation{Universidad Michoacana de San Nicolás de Hidalgo, Morelia, Mexico }

\author[0000-0002-9790-1299]{J.A.~Goodman}
\affiliation{Department of Physics, University of Maryland, College Park, MD, USA }

\author{S.~Groetsch}
\affiliation{Department of Physics, Michigan Technological University, Houghton, MI, USA }

\author{J.~Gyeong}
\affiliation{Department of Physics, Sungkyunkwan University, Suwon 16419, South Korea}

\author[0000-0001-9844-2648]{J.P.~Harding}
\affiliation{Los Alamos National Laboratory, Los Alamos, NM, USA }

\author[0000-0001-5169-723X]{I.~Herzog}
\affiliation{Department of Physics and Astronomy, Michigan State University, East Lansing, MI, USA }

\author[0000-0002-5447-1786]{D.~Huang}
\affiliation{Department of Physics and Astronomy, University of Delaware, Newark, DE, USA}

\author[0000-0002-5527-7141]{F.~Hueyotl-Zahuantitla}
\affiliation{Universidad Autónoma de Chiapas, Tuxtla Gutiérrez, Chiapas, México}

\author[0000-0002-3302-7897]{P.~Hüntemeyer}
\affiliation{Department of Physics, Michigan Technological University, Houghton, MI, USA }

\author[0000-0001-5811-5167]{A.~Iriarte}
\affiliation{Instituto de Astronom\'{i}a, Universidad Nacional Autónoma de México, Ciudad de México, México.}

\author{S.~Kaufmann}
\affiliation{Universidad Politecnica de Pachuca, Pachuca, Hgo, Mexico }

\author[0000-0003-4785-0101]{D.~Kieda}
\affiliation{Department of Physics and Astronomy, University of Utah, Salt Lake City, UT, USA }

\author{K.~Leavitt}
\affiliation{Department of Physics, Michigan Technological University, Houghton, MI, USA }

\author[0000-0002-2153-1519]{J.S.H.~Lee}
\affiliation{Department of Physics, Yonsei University, Seoul, South Korea}

\author[0000-0001-5516-4975]{H.~León Vargas}
\affiliation{Instituto de F\'{i}sica, Universidad Nacional Autónoma de México, Ciudad de Mexico, Mexico }

\author[0000-0001-8825-3624]{A.L.~Longinotti}
\affiliation{Instituto de Astronom\'{i}a, Universidad Nacional Autónoma de México, Ciudad de México, México.}

\author[0000-0003-2810-4867]{G.~Luis-Raya}
\affiliation{Universidad Politecnica de Pachuca, Pachuca, Hgo, Mexico }

\author[0000-0001-8088-400X]{K.~Malone}
\affiliation{Los Alamos National Laboratory, Los Alamos, NM, USA }

\author[0000-0001-9052-856X]{O.~Martinez}
\affiliation{Facultad de Ciencias F\'{i}sico Matemáticas, Benemérita Universidad Autónoma de Puebla, Puebla, Mexico }

\author[0000-0002-2824-3544]{J.~Martínez-Castro}
\affiliation{Centro de Investigaci\'on en Computaci\'on, Instituto Polit\'ecnico Nacional, M\'exico City, M\'exico.}

\author[0000-0001-7714-0704]{H.~Martínez-Huerta}
\affiliation{Departamento de F\'{i}sica y Matem\'{a}ticas, Universidad de Monterrey, Av. Morones Prieto 4500, San Pedro Garza Garc\'{i}a 66238, Nuevo Le\'{o}n, M\'{e}xico}

\author[0000-0002-2610-863X]{J.A.~Matthews}
\affiliation{Dept of Physics and Astronomy, University of New Mexico, Albuquerque, NM, USA }

\author[0000-0002-8390-9011]{P.~Miranda-Romagnoli}
\affiliation{Universidad Autónoma del Estado de Hidalgo, Pachuca, Mexico }

\author{P.E.~Mirón-Enriquez}
\affiliation{Instituto de Astronom\'{i}a, Universidad Nacional Autónoma de México, Ciudad de México, México.}

\author[0000-0001-9361-0147]{J.A.~Morales-Soto}
\affiliation{Universidad Michoacana de San Nicolás de Hidalgo, Morelia, Mexico }

\author[0000-0002-1114-2640]{E.~Moreno}
\affiliation{Facultad de Ciencias F\'{i}sico Matemáticas, Benemérita Universidad Autónoma de Puebla, Puebla, Mexico }

\author[0000-0002-7675-4656]{M.~Mostafá}
\affiliation{School of Science \& Engineering, University of Missouri--Kansas City, Kansas City, MO, USA}

\author{M.~Najafi}
\affiliation{Department of Physics, Michigan Technological University, Houghton, MI, USA }

\author[0000-0003-1059-8731]{L.~Nellen}
\affiliation{Instituto de Ciencias Nucleares, Universidad Nacional Autónoma de Mexico, Ciudad de Mexico, Mexico }

\author{Y.~Niroula}
\affiliation{Temple University, Department of Physics, 1925 N. 12th Street, Philadelphia, PA 19122, USA}

\author[0000-0002-6859-3944]{M. U. Nisa}
\affiliation{Dept. of Physics and Astronomy, Michigan State University, East Lansing, MI 48824, USA}

\author[0000-0001-7099-108X]{R.~Noriega-Papaqui}
\affiliation{Universidad Autónoma del Estado de Hidalgo, Pachuca, Mexico }

\author[0009-0009-2481-6921]{M.~Osorio-Archila}
\affiliation{Instituto de Astronom\'{i}a, Universidad Nacional Autónoma de México, Ciudad de México, México.}

\author{E.~Ponce}
\affiliation{Facultad de Ciencias F\'{i}sico Matemáticas, Benemérita Universidad Autónoma de Puebla, Puebla, Mexico }

\author[0000-0002-8774-8147]{Y.~Pérez Araujo}
\affiliation{Instituto de F\'{i}sica, Universidad Nacional Autónoma de México, Ciudad de Mexico, Mexico }

\author[0000-0001-5998-4938]{E.G.~Pérez-Pérez}
\affiliation{Universidad Politecnica de Pachuca, Pachuca, Hgo, Mexico }

\author[0000-0003-1327-0838]{D.~Rosa-González}
\affiliation{Instituto Nacional de Astrof\'{i}sica, Óptica y Electrónica, Puebla, Mexico }

\author[0000-0002-4204-5026]{M.~Roth}
\affiliation{Los Alamos National Laboratory, Los Alamos, NM, USA }

\author[0000-0001-6079-2722]{A.~Sandoval}
\affiliation{Instituto de F\'{i}sica, Universidad Nacional Autónoma de México, Ciudad de Mexico, Mexico }

\author[0000-0001-8644-4734]{M.~Schneider}
\affiliation{Department of Physics, University of Maryland, College Park, MD, USA }

\author{J.~Serna-Franco}
\affiliation{Instituto de F\'{i}sica, Universidad Nacional Autónoma de México, Ciudad de Mexico, Mexico }

\author{S.~Shankar}
\affiliation{Temple University, Department of Physics, 1925 N. 12th Street, Philadelphia, PA 19122, USA}

\author[0000-0002-1012-0431]{A.J.~Smith}
\affiliation{Department of Physics, University of Maryland, College Park, MD, USA }

\author{Y.~Son}
\affiliation{Department of Physics, Sungkyunkwan University, Suwon 16419, South Korea}

\author[0000-0002-1492-0380]{R.W.~Springer}
\affiliation{Department of Physics and Astronomy, University of Utah, Salt Lake City, UT, USA }

\author{O.~Tibolla}
\affiliation{Universidad Politecnica de Pachuca, Pachuca, Hgo, Mexico }

\author[0000-0001-9725-1479]{K.~Tollefson}
\affiliation{Department of Physics and Astronomy, Michigan State University, East Lansing, MI, USA }

\author[0000-0002-1689-3945]{I.~Torres}
\affiliation{Instituto Nacional de Astrof\'{i}sica, Óptica y Electrónica, Puebla, Mexico }

\author[0000-0002-7102-3352]{R.~Torres-Escobedo}
\affiliation{Tsung-Dao Lee Institute \& School of Physics and Astronomy, Shanghai Jiao Tong University, 800 Dongchuan Rd, Shanghai, SH 200240, China}

\author[0000-0003-0715-7513]{E.~Varela}
\affiliation{Facultad de Ciencias F\'{i}sico Matemáticas, Benemérita Universidad Autónoma de Puebla, Puebla, Mexico }

\author[0000-0001-6876-2800]{L.~Villaseñor}
\affiliation{Facultad de Ciencias F\'{i}sico Matemáticas, Benemérita Universidad Autónoma de Puebla, Puebla, Mexico }

\author[0009-0005-7243-1402]{H.~Wu}
\affiliation{Dept. of Physics and Wisconsin IceCube Particle Astrophysics Center, University of Wisconsin{\textemdash}Madison, Madison, WI, USA}

\author[0000-0001-6798-353X]{X.~Wang}
\affiliation{Department of Physics, Missouri University of Science and Technology, Rolla, MO, US}

\author{Z.~Wang}
\affiliation{Department of Physics, Missouri University of Science and Technology, Rolla, MO, US}

\author[0000-0003-2141-3413]{I.J.~Watson}
\affiliation{University of Seoul, Seoul, Rep. of Korea}

\author[0009-0006-3520-3993]{S.~Yu}
\affiliation{Department of Physics, Pennsylvania State University, University Park, PA, USA}

\author[0000-0002-9307-0133]{S.~Yun-Cárcamo}
\affiliation{Dept. of Physics, Drexel University, 3141 Chestnut Street, Philadelphia, PA 19104, USA
}

\author{X.~Zhang}
\affiliation{Institute of Nuclear Physics Polish Academy of Sciences, PL-31342 IFJ-PAN, Krakow, Poland }

\author[0000-0003-0513-3841]{H.~Zhou}
\affiliation{Tsung-Dao Lee Institute \& School of Physics and Astronomy, Shanghai Jiao Tong University, 800 Dongchuan Rd, Shanghai, SH 200240, China}

%% file: IceCube-authors.tex
\author[0000-0001-6141-4205]{R. Abbasi}
\affiliation{Department of Physics, Loyola University Chicago, Chicago, IL 60660, USA}

\author[0000-0001-8952-588X]{M. Ackermann}
\affiliation{Deutsches Elektronen-Synchrotron DESY, Platanenallee 6, D-15738 Zeuthen, Germany}

\author{J. Adams}
\affiliation{Dept. of Physics and Astronomy, University of Canterbury, Private Bag 4800, Christchurch, New Zealand}

\author[0000-0003-2252-9514]{J. A. Aguilar}
\affiliation{Universit{\'e} Libre de Bruxelles, Science Faculty CP230, B-1050 Brussels, Belgium}

\author[0000-0003-0709-5631]{M. Ahlers}
\affiliation{Niels Bohr Institute, University of Copenhagen, DK-2100 Copenhagen, Denmark}

\author[0000-0002-9534-9189]{J.M. Alameddine}
\affiliation{Dept. of Physics, TU Dortmund University, D-44221 Dortmund, Germany}

\author[0009-0001-2444-4162]{S. Ali}
\affiliation{Dept. of Physics and Astronomy, University of Kansas, Lawrence, KS 66045, USA}

\author{N. M. Amin}
\affiliation{Bartol Research Institute and Dept. of Physics and Astronomy, University of Delaware, Newark, DE 19716, USA}

\author[0000-0001-9394-0007]{K. Andeen}
\affiliation{Department of Physics, Marquette University, Milwaukee, WI 53201, USA}

\author[0000-0003-4186-4182]{C. Arg{\"u}elles}
\affiliation{Department of Physics and Laboratory for Particle Physics and Cosmology, Harvard University, Cambridge, MA 02138, USA}

% ...
